\documentclass[preprint,12pt]{elsarticle}

\usepackage{graphicx}
\usepackage{amssymb}
\usepackage{amsmath}
\usepackage{animate}
\usepackage{xcolor}
\usepackage{comment}
\usepackage{fancyhdr}
\usepackage[hyphens]{url}

\usepackage{lineno}
\usepackage{subcaption}

\makeatletter
\def\ps@pprintTitle{
  \def\@oddfoot{\mycopyrightnotice}
  \def\@evenfoot{}
}
\def\mycopyrightnotice{
  {\footnotesize
  \begin{minipage}{\textwidth}
  Preprint submitted to Renewable and Sustainable Energy Reviews\\
  Copyright~\copyright~2019. Licensed under the Creative Commons CC-BY-NC-ND
  \end{minipage}
  }
}

\journal{Renewable and Sustainable Energy Reviews}

\begin{document}

\begin{frontmatter}




\title{Fleetwide data-enabled reliability improvement of wind turbines}

\cortext[cor1]{Corresponding author}

\author[ai]{Timothy Verstraeten\corref{cor1}}
\ead{tiverstr@vub.be}
\author[ai]{Ann Now\'e}
\author[nrel]{Jonathan Keller}
\author[nrel]{Yi Guo}
\author[nrel]{Shuangwen Sheng}
\author[avrg,owi]{Jan Helsen\corref{cor1}}
\ead{jahelsen@vub.be}

\address[ai]{Artificial Intelligence Lab, Vrije Universiteit Brussel, Pleinlaan 2, 1050 Brussels, Belgium}
\address[avrg]{Acoustics \& Vibrations Research Group, Vrije Universiteit Brussel, Pleinlaan 2, 1050 Brussels, Belgium}
\address[owi]{OWI-Lab, Celestijnenlaan 300b, 3001 Heverlee, Belgium}
\address[nrel]{National Renewable Energy Laboratory, National Wind Technology Center, Golden, CO, USA}

\begin{abstract}
Wind farms are an indispensable driver toward renewable and nonpolluting energy resources. However, as ideal sites are limited, placement in remote and challenging locations results in higher logistics costs and lower average wind speeds. Therefore, it is critical to increase the reliability of the turbines to reduce maintenance costs. Robust implementation requires a thorough understanding of the loads subject to the turbine's control. Yet, such dynamically changing multidimensional loads are uncommon with other machinery, and generally underresearched. Therefore, a multitiered approach is proposed to investigate the load spectrum occurring in wind farms. Our approach relies on both fundamental research using controllable test rigs, as well as analyses of real-world loading conditions in high-frequency supervisory control and data acquisition data. A method is introduced to detect operational zones in wind farm data and link them with load distributions. Additionally, while focused research further investigates the load spectrum, a method is proposed that continuously optimizes the farm's control protocols without the need to fully understand the loads that occur. A case of gearbox failure is investigated based on a vast body of past experiments and suspect loads are identified. Starting from this evidence on the cause and effects of dynamic loads, the potential of our methods is shown by analyzing real-world farm loading conditions on a steady-state case of wake and developing a preventive row-based control protocol for a case of cascading emergency brakes induced by a storm.
\end{abstract}


\end{frontmatter}

\section{Wind as a cost-effective renewable energy source}

As a society, we recognize the effects of climate change \cite{climate_change_2007, climate_change_2018}. Moreover, in Europe there is a drive to close nuclear power plants \cite{closing_nuclear_plants_2012}. This reality presents an opportunity to strategically increase the percentage of renewable power generation. In 2017, the global installed wind capacity increased by 10\% \cite{global_wind_2018}. Offshore wind is expected to play a large role, experiencing a growth of 101\% in European installations in 2017 compared to 2016 \cite{offshore_wind_2018}. A major hurdle to the development and acceptance of renewable electricity sources is ensuring that they are cost competitive with fossil-fuel generation. More specifically, they must meet or exceed the ratio between expected production and lifetime costs often defined as levelized cost of energy (LCOE) \cite{lcoe_2005}.

For long-term viability, offshore wind must significantly improve its cost efficiency \cite{offshore_efficiency_2017}. Today, land-based wind energy is cost competitive at wind sites with a strong resource \cite{onshore_wind_sites_2015}; however, in Europe ideal onshore sites are already becoming populated, so expansion of wind farms will require placement in increasingly more remote and challenging locations, resulting in higher logistics costs and lower average wind speeds \cite{cost_analysis_2011, cost_analysis_2017}. To reduce LCOE for wind energy, reliability of turbines must rapidly be improved, even in harsh environments, and reduce operations and maintenance costs in general \cite{lcoe_2018}.

In this work, reliability is considered to be the likelihood that a turbine and its components will meet its prescribed design life \cite{design_life_1994}. In that sense, an unreliable turbine or component fails because of design flaws caused by the fact that the design does not accurately capture its environmental loading conditions.
A robust machine is one optimized to perform its function in any condition that it may encounter during its lifetime, even within a wide range of environmental load cases. Overdesigning is not a viable solution, as the capital cost of the turbine will increase too drastically. Truly robust implementation at the lowest cost features minimum possible safety factors as the loads are so intimately understood. Therefore, to improve reliability, a study must be performed to thoroughly understand the spectrum of loads that occur during turbine operations subject to its control behavior in all operating environments likely to be encountered, and then systematically incorporate them in the design loads.

The case of dynamically changing multidimensional loads of such great magnitude as found in wind turbines is uncommon with other industrial machinery, and generally underresearched \cite{loads_under_researched_2015, load_failure_2017}. The loading conditions leading to failure are often a combination of dynamics (i.e., rapidly fluctuating wind speeds) compounded by grid-based, electrically originating loading conditions \cite{load_causes_2016}. Site-specific conditions and corresponding turbine responses are expected to play a significant role, because the most significant failures occur only on a subset of the complete \emph{fleet} of turbines \cite{failures_subset_2016}.
Sites are differential in wind resource, terrain, seasonal events, and atmospheric changes. Additionally, the electricity grid has different subregions, each with specific utility interconnection requirements, and even fluctuations in behavior within the region \cite{grid_instability_2007}. This poses significantly fluctuating loading conditions on the turbine that are, as this work shows, at the origin of failure.

Although it is tempting to consider a wind turbine as a generic entity, and impose operational control from this perspective, in reality each individual mechanical unit is unique. Therefore, maximum reliability can best be achieved through the use of the methodology described herein to allow treating the fleet as a data-compiling collective, and capturing the similarities between turbines that exist on a statistical level, while acknowledging the uniqueness in their specific operational behavior. By investigating the links between operational behavior and loads, each turbine can learn optimized responses to avoid loading conditions that may lead to failure.

Although wind farm control research mainly focuses on static loads and power production \cite{overview_farm_control_2014, tutorial_control_2017}, the reduction of dynamic loads through operational measures has received less attention. However, recent evidence suggests that dynamic loads induce failure \cite{load_failure_2017, slipping_2015, load_failure_2018, drivetrain_investigation_2018, drivetrain_investigation_2016}. Preventing failures has a direct impact on the availability of the turbines and it reduces lifetime costs \cite{lcoe_2018}. Therefore, this work proposes a multitiered framework to investigate and understand the link between failures and dynamic loads, such that preventive measures can be implemented through targeted wind farm control strategies.

\section{Strategy and areas of focus}

\begin{figure}[!hb]
\begin{center}
\includegraphics[scale=0.4]{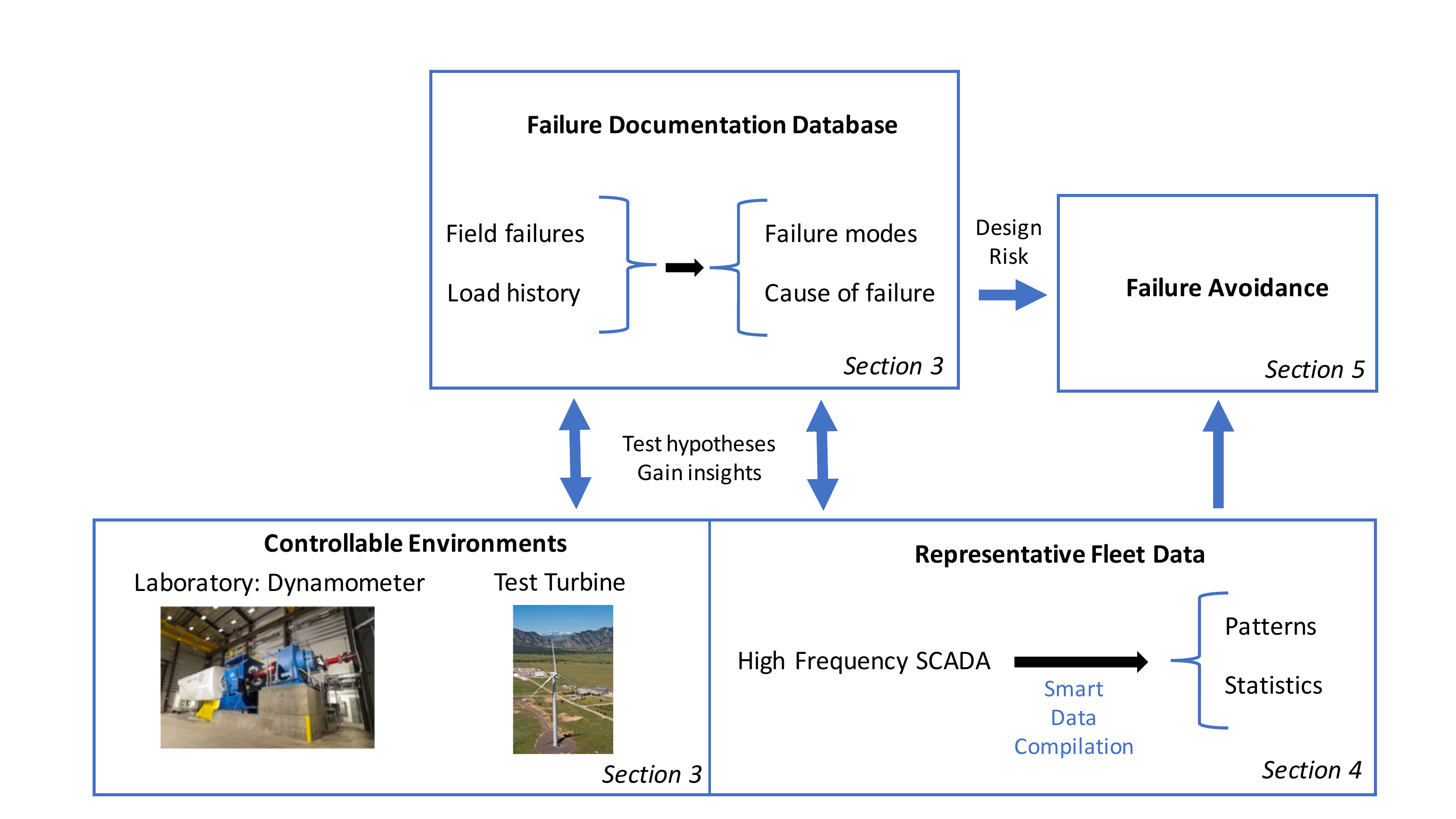}
\caption{Multitiered investigation strategy for optimizing wind turbine reliability. Photos by Dennis Schroeder, NREL 31778 and 21884}
\label{fig:overview}
\end{center}
\end{figure}

In order to understand the load cases and the root of their failure mechanisms, a multipronged investigation that draws on all data sources is proposed, depicted in Figure~\ref{fig:overview}. Central to our research is a database of documented failures \cite{grd_2016}. Gaining insight into the causes of the failures and associated loads provides the necessary knowledge for optimizing turbine design.
Naturally, when a failure occurs in the field, the origin of the failure has to be investigated. Our proposed approach starts from the failure modes and relies on both traditional experimentation in controllable test rigs, as well as analyses of trends occurring in the real-world wind farm.

Starting from historic data, investigations are based on trends occurring in the field and focus on suspect loads and mechanisms that are known to be prevalent. The gearbox subsystem of the drivetrain has historically constituted a disproportionately high level of failures in wind turbines \cite{load_causes_2016, gearbox_failure_causes_2009}. The issues with reliability of this subsystem are responsible for a significant part of the unrealized design life goals. In order to effectively and rapidly increase the reliability of these systems, the focus is on critical components; those that can cause long downtimes, occur in many turbines, or result in high component replacement costs. Detailed inspection reports on gearbox failure modes are the starting point. The database encompasses 40\% of the U.S. land-based fleet and a subset consisting of 280 cases with heavily detailed accounts of gearbox failure modes is examined \cite{gearbox_failure_modes_2015}. For the gearbox, two specific failed components are focused on. Namely, the high-speed shaft (HSS) bearing, which is responsible for the largest percentage (48\%) of failures of any subcomponent. Although relatively inexpensive to replace, the sheer number of these failures has been problematic. Additionally, the planetary bearing, which causes only 6.6\% of failures, requires removal of the gearbox from the turbine with a large crane, leading to excessive cost and downtime \cite{bearing_failure_2016}. Controlled lab experiments are performed to investigate two cases of failure: planetary bearing failure resulting from nontorque wind loading and HSS bearing failure caused by dynamic grid events.

Although gearbox failure modes can be investigated in controlled environments, the loading conditions induced during these experiments are only relevant when they are representative for real-world wind farm sites.
Therefore, another element of our research approach is introduced, involving the compilation and analysis of 1-second field data that is provided by a supervisory control and data acquisition system (SCADA), in order to pinpoint the static loading conditions occurring in the wind farm. By identifying zones of operationally similar turbines and building zone-specific load profiles based on field data, relevant patterns and statistics can be extracted and communicated to the designer. This process enables a detailed understanding of the links between operating conditions and loads occurring in the farm.

Finally, while focused research investigates the links between loading conditions and failure modes, the reliability of the fleet is optimized in our last step by introducing a data-driven controller that creates failure avoidance strategies at the wind farm level. 
By optimizing beyond established control protocols, fleet productivity is increased and failures are prevented during dynamic events. This capability is directly related to the LCOE, as it improves productivity and reduces lifetime costs. The interaction between the control designer and data-driven controller is important. The designer composes the format and objective of the controller in terms of failure and productivity based on past research on loading conditions and failure modes, whereas static load profiles guide the initiation of the controller. More specifically, discrepancies in the load profiles---together with other data sources---can identify which event is taking place; therefore, the appropriate control strategy can be initiated. 

Recent studies promote the use of high-frequency SCADA data for analyzing turbine dynamics \cite{high_freq_scada_2017, high_freq_scada_2019}. Using high-frequency data allows for deeper and detailed insights into load histories. Moreover, implementing preventive measures during brief failure-driven events may be impossible at lower resolutions. An example of this is given in Section~\ref{sec:failure_avoidance}, in which preventive actions must be taken rapidly.

Currently, the sharing of field data for research is highly limited \cite{share_data_2016}. Nonetheless, the utility of our techniques is demonstrated based on various focused study cases. The need for our data-driven, physics-based investigation strategy is shown in Section~\ref{sec:gearbox} by providing evidence for dynamic loads that cause gearbox failure, based on 10 years of dedicated testing and field wind turbine monitoring in the context of the Gearbox Reliability Collaborative (GRC) \cite{grc_2011}. In Section~\ref{sec:fleet_loads}, a mechanism based on 1-second SCADA field data is presented to extract operational zones and link them to corresponding static load profiles.
Finally, in Section~\ref{sec:failure_avoidance}, a reinforcement learning mechanism is introduced for optimizing a parametric row-based control policy to prevent loads induced by emergency brakes under excessive wind speed events.

\section{Related work}
\label{sec:related}

Previous work on wind farm control strategies has focused on maximizing power production and minimizing fatigue loads \cite{overview_farm_control_2014, tutorial_control_2017}. A recent well-studied approach is to mitigate the wake effect by using yaw and tilt control on the individual turbines to optimize both criteria for the complete farm through field data \cite{wake_mitigation_2014, yaw_tilt_control_2018, wake_mitigation_2018}. Optimizing the yaw and tilt of upstream turbines can deflect wake, which allows downstream turbines to extract more energy. However, this work differentiates itself by focusing on dynamic loads and how they are linked to events. It demonstrates how insights on event-driven dynamic loads can be incorporated directly in farm control strategies to prevent failure.

Investigating the cause and effects between dynamic loads and failure modes of gearboxes has recently gained popularity. Evidence suggests that blade-pitch faults induce axial forces on the main bearing of the drivetrain \cite{drivetrain_investigation_2016}. Other work shows the relationship between dynamic loads and electrical components, with a specific focus on power converter faults \cite{drivetrain_investigation_2018}. Dynamic events, such as grid loss and emergency stops, also heavily impact the lifetime of the turbine components \cite{jiang2014dynamic}. This work discusses lessons learned from analyzing the potential of bearing slip during grid loss events. Results and insights on dynamic loads are analyzed and interpreted based on vast corpus of systematic realistic experiments on the GRC drivetrain \cite{grc_phases_12_2011, grc_phase_3}.

\section{Evidence for dynamic loads causing reduced bearing life}
\label{sec:gearbox}

\begin{figure}[!hb]
\begin{center}
\includegraphics[scale=0.8]{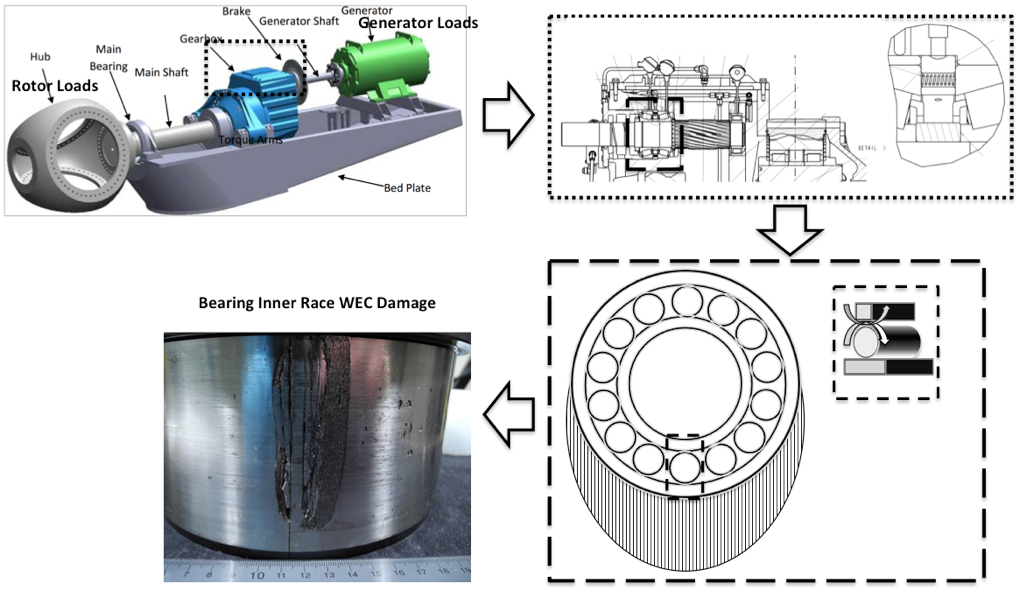}
\caption{Clockwise from top-left, illustration of rotor, generator and high speed stage bearing locations; typical bearing load zone; bearing rolling element and example of bearing WEC failure \cite{wec_example_2016}.}
\label{fig:wec_example}
\end{center}
\end{figure}

Laboratory testing using dedicated dynamometer setups \cite{grc_phase_3} shows the link between unfavorable (dynamic) loads and bearing failures in the field \cite{grc_phases_12_2011, bearing_slip_2018}. These test results show that a large majority (71\%) of the HSS bearing failures appears to be caused by axial cracking resulting from local microstructural changes. More specifically, a phenomenon known as white etch cracking (WEC) was observed frequently \cite{wec_example_2016}. WEC owes its name to the appearance of white regions when etched with Nital \cite{wec_etches_2006}. Figure~\ref{fig:wec_example} shows the location of the bearing and an example of WEC. Bearings are expected to fail because of high cycle fatigue near their probabilistic design life. However, it has been shown that failure caused by WEC can result in a reduction of as much as 95\% of the design life \cite{slip_emergency_evidence_2013}. Load states that have been linked to WEC arise from loading zones not anticipated in design. Both industry and academic research is ongoing to unveil the failure mechanism \cite{wec_2018}.

In normal operation, all bearing roller trajectories should be close to the kinematic optimal rolling path. Deviation from the normal trajectory indicates slipping of the roller. Figure~\ref{fig:wec_example} shows the traditional load zone for a cylindrical roller bearing. Rollers are continuously accelerating and decelerating when they pass through the load zone. Excessive slipping is induced when acceleration makes kinematic pure rolling impossible. Excessive slip is expected to damage bearings because of sliding effects between the roller and raceway, causing excessive wear and potentially crack initiation. It has been shown in literature that this slipping can be induced by dynamic events \cite{slipping_2015}, but with only limited demonstration on a full-scale test article. Specifically, roller slip induced by dynamic load cases is investigated \cite{grid_loss_2016, system_response_2016, bearing_slip_2018}.
Dynamic loads can be induced by certain control actions or erroneous turbine responses, such as blade pitching, startup, emergency braking or yawing \cite{responses_2009}, during dynamic wind or grid events.

It is necessary to investigate the field activity that creates suspect loading zones and, in parallel, study the failure mechanism in the laboratory, as shown in Figure~\ref{fig:overview}. Controlled dynamometer experiments allow fundamental mechanisms to be studied. However, results are only valid if the input parameters represent real load states. Therefore, to develop true suspect loading profiles, a test turbine is monitored with extensive instrumentation \cite{grc_2011}. In this work, turbulent wind conditions and turbine errors that triggered emergency stop events were reported until failure happened \cite{grc_desc_loading_2011}. The loading signatures obtained from the field measurements during emergency stop events were used as a design input for the load cases to be tested on the dynamometer. This type of instrumentation, and corresponding data analysis, cannot feasibly be implemented fleet-wide as it is cost prohibitive. Combining field data with laboratory analysis allows for the development of design load data that is far more congruent with what occurs in a turbine’s actual environment.

Experiments were conducted on the GRC drivetrain mounted on the National Renewable Energy Laboratory dynamometer \cite{grc_2011}, shown in Figure~\ref{fig:overview}. This gearbox is a good test article given that its design is in public domain and its comprehensive instrumentation package \cite{gearbox_instrumentation_2013}. The dynamometer has 4 degrees of freedom for the load input control to represent the complex flow patterns of the wind. Additionally, control of the local electrical grid is possible to simulate grid instability events.

Two fault scenarios were investigated. First, the triggering of roller slip in the HSS linked to WEC. Grid-loss events were applied to investigate these conditions \cite{guo2015planetary, lacava2013three}. Second, planetary bearing failures were investigated \cite{grc_phases_12_2011}.

To test the effects of grid instability, the drivetrain was subjected to several worst-case grid-loss events initiated at different power levels \cite{grid_loss_2016}. During these events, the generator currents dropped to zero instantaneously. The system response was consistent and characterized by torque reversals driven by a main drivetrain resonance linked to the rotor and generator inertias connected by the drivetrain flexibility \cite{system_response_2016}. These torque reversals result in multidimensional loading at the HSS, thereby causing the rollers to slip excessively. Because these slip conditions increase the risk of bearing damage, it can be concluded that an electric event has the potential to cause unfavorable dynamic mechanical loading and possible degradation of the gearbox system.

The second main failure type for gearboxes was planetary bearing failure. For these bearings, overloading conditions are important drivers for degradation. Rotor moments significantly contribute to these failures and they are insufficiently understood. These moments are caused by rotor-overhung weight, excessive wind shear, yawing out of the direction of the wind, and dynamic loading induced during certain turbine control actions, such as blade pitching or emergency braking, particularly when the controller experiences an error in the performed action. These conditions lead to unfavorable wind loading on the turbine blades that, in many turbine designs, is subsequently passed through the drivetrain to the nacelle \cite{drivetrain_loads_2014}.

Experimentally measured main shaft loads of the field wind turbine were used as load inputs on the dynamometer. The inability of the planet carrier bearings to counteract the main shaft bending loads in combination with gear misalignment causes overloading of one of the two planet bearings and hence significantly accelerates planet bearing failure \cite{grc_2011, bearing_overload_2012}. These tests support the proposed hypothesis that the response to dynamic wind events can initiate mechanical degradation of planetary bearings.


\section{Load profiling on operational field data}
\label{sec:fleet_loads}

To obtain statistically relevant insights into real-world loading conditions occurring in a wind farm, another step in the research approach is proposed (see Figure~\ref{fig:overview}), in which operational 1-second SCADA field data is analyzed. By identifying patterns in operational fleet data and connecting them with fundamental research, as described earlier, the designer gains knowledge about the loading conditions present in the wind farm for various environmental contexts. This knowledge can be incorporated in the failure avoidance control mechanism. More specifically, deviations from the expected loading conditions can be detected and an appropriate controller strategy can be initiated to adequately deal with this deviation.

As shown in this work, loading conditions can be directly related to operational \emph{zones} within the wind farm. In order to extract these zones, operational fleet data are modeled using Bayesian Gaussian mixture models (GMMs) \cite{bayes_gmm_2006}. These models can be used as a clustering technique, as they are able to derive several Gaussian distributions from data (hence the name). As each cluster is described using a Gaussian, the parameters that should be learned are the mean vectors and covariance matrices. The Bayesian variant introduces a Dirichlet process prior \cite{bayes_gmm_2006} to infer the number of Gaussians used to model the data. This reduces model complexity to maintain interpretability and to prevent overfitting.

Restricting the data to a certain distributional form gives GMMs several desirable properties: the automatic inference of the number of clusters, the robust differentiation between noise and outliers, and the ability to perform well on small data sets. This provides a suitable mechanism to detect the frequently changing patterns in fleet control data, and to detect similarities between the turbines' behaviors for specific environmental loading conditions.

Many other clustering algorithms exist  \cite{clustering_overview}. K-means is one of the most frequently used techniques \cite{k_means_1967}, because of its simplicity and efficiency \cite{k_means_2010}. It aims to establish a set of centroids and assigns each data point to the closest centroid. Although simple to use and useful to gain early insights, it performs poorly with noisy data \cite{k_means_2010}---which is often the case with distance-based clustering methods \cite{clustering_overview}. GMMs inherently model this noise in the data, which provides a suitable mechanism for the noisy outliers that can occur in high-frequency SCADA data. Another type of clustering that could be suitable for our setting is fuzzy clustering \cite{fuzzy_clustering_1989}. Although highly similar to GMMs, they are harder to interpret and therefore lack theoretical grounding with respect to given data compared to GMMs \cite{gmm_1988}.

Once the operational zones have been identified, statistically relevant insights into farm dynamics can be extracted. For this purpose, the concept of \emph{load profiles} is introduced, which contain statistical properties about the loads of the turbines within a particular operational zone. More specifically, it is a mapping from operational parameters (e.g., power production and rotor speed) to load statistics for a particular operational zone. For example, productivity within the farm can be decomposed into zones within the farm, and load profiles can be computed per zone. Such statistics are useful for the failure avoidance controller, as it describes which turbines should observe similar load spectra for specific environmental conditions (see Section~\ref{sec:failure_avoidance}).


The benefits of load profiles are investigated using 3 months of 1-second SCADA data, measured over a farm comprised of 55 wind turbines. As the operational zones and load profiles are dependent on the environmental conditions in the farm, the incoming average wind vector is reported for each profile. To reduce measurement noise, the average of each SCADA parameter is computed over 2-min windows. Based on a qualitative analysis of the full data set, 2 min are considered appropriate to obtain consistent results, while maintaining the finer trends in the operational dynamics.

A steady-state case is chosen from the analysis to illustrate the proposed method for which the environmental conditions are stable.\footnote{The data for turbine 10/4 are not available.}
The stability is ensured as the incoming wind vector, measured by averaging the SCADA wind speed over all turbines per second, does not change significantly during the 2-minute window. This process is verified by reporting the minimum and maximum wind vectors for 10 minutes before and during the 2-minute window, which are $7.9$ m/s at $235.0^{\circ}$ and $8.7$ m/s at $235.7^{\circ}$, respectively. The average incoming wind vector at the farm site is at $235.4^{\circ}$ and $8.2$ m/s, which is computed on the SCADA data of all turbines. At this wind speed, the turbines are mostly operating on torque control. Rotor speed (rpm) and power production (kW) describe the generator torque \cite{torque_control_2004}, which means they are sufficient parameters to describe the operational regime. As the rotor speed only takes into account \emph{effective} wind speed, from which power is generated, the rotor speed provides a more reliable estimate to describe operating conditions than standard wind speed measurements \cite{effective_wind_speed_2015}. Per turbine, the averages of these operational parameters are computed---resulting in a single data point per turbine---and the clustering method is applied. The Bayesian GMMs are optimized using variational inference \cite{bayes_gmm_2006} until convergence (error $ < 10^{-5}$). The Dirichlet process prior is chosen based on a qualitative analysis on the 3-month SCADA data. The parametrization of the Dirichlet process prior is in terms of the expected maximum amount of clusters. The expected maximum amount is set to 6, as it yields the most consistent and reliable results over the 3-month SCADA data set.

\begin{figure}[!hb]
\begin{subfigure}{0.48\textwidth}
\begin{center}
\includegraphics[scale=0.47,trim=1cm 1.3cm 0cm 0cm]{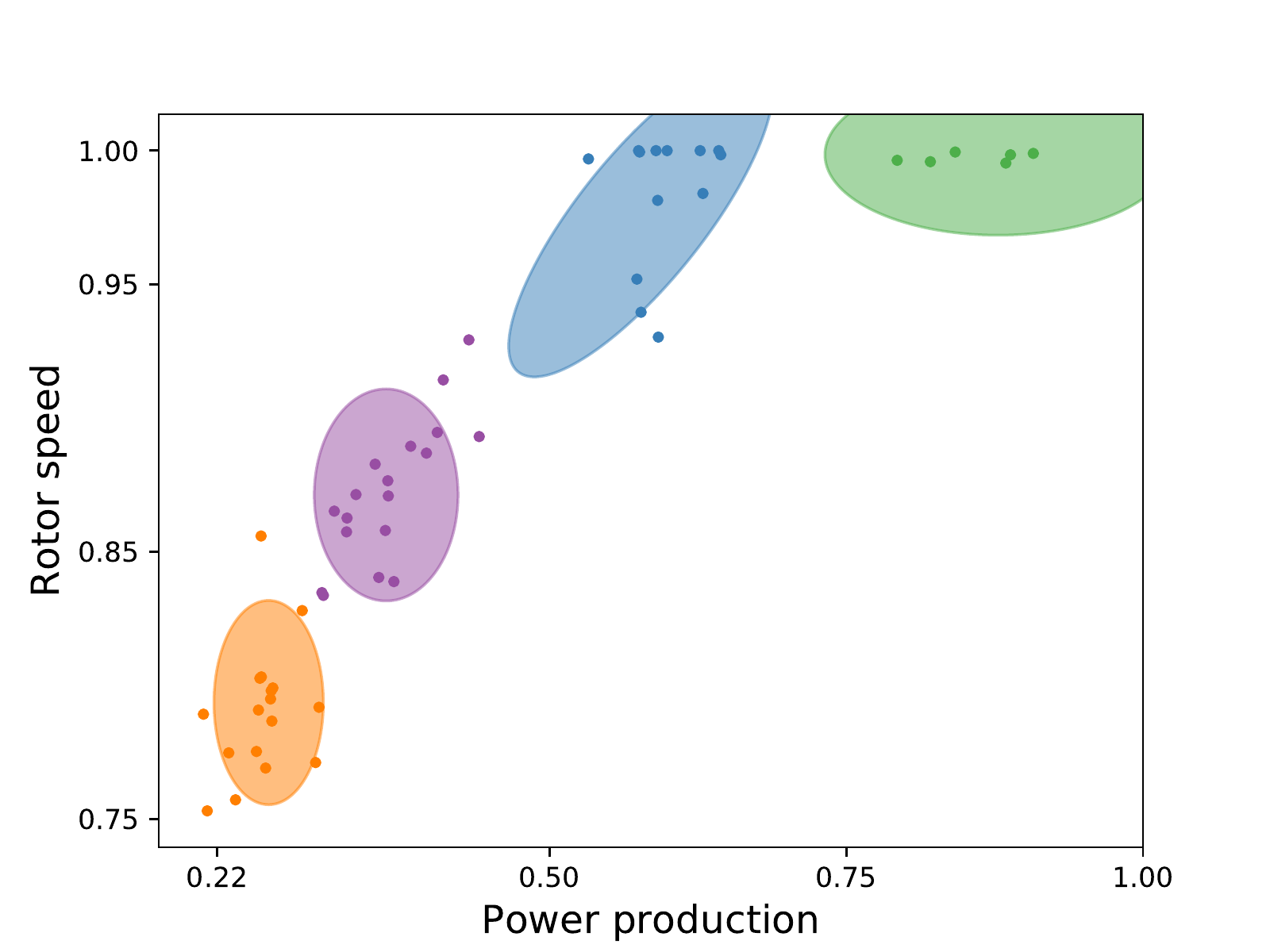}
\end{center}
\caption{Operational zones}
\end{subfigure}
\begin{subfigure}{0.48\textwidth}
\begin{center}
\includegraphics[scale=0.47,trim=0cm 1.3cm 0cm 0cm]{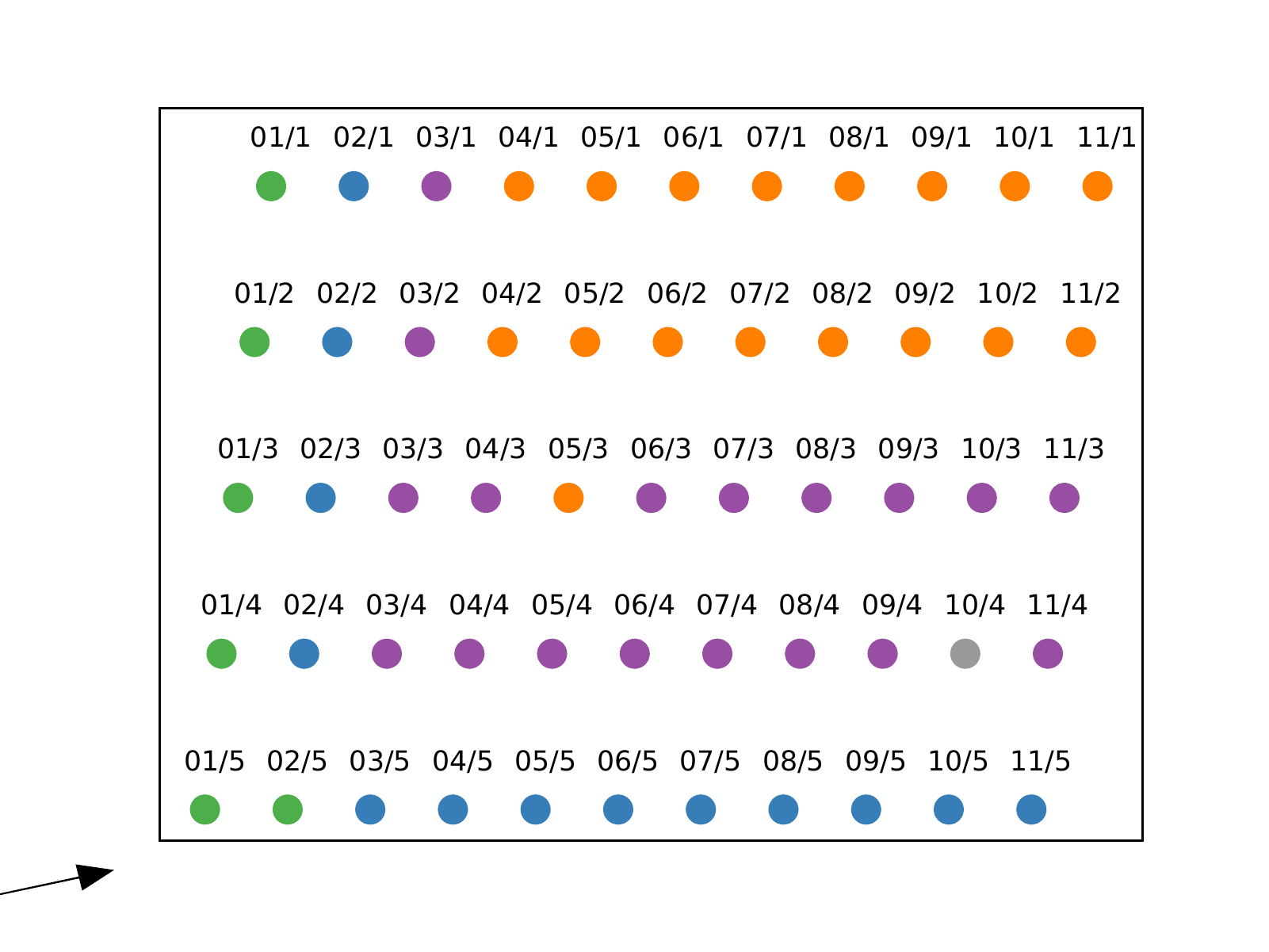}
\end{center}
\caption{Wind farm view}
\end{subfigure}
\caption{The operational zones identified through GMM clustering: (a) Each measurement (point) is assigned to one 2D Gaussian (ellipse). Each axis is normalized between 0 and 1, respectively, denoting the minimum and maximum value possible. In the wind farm, a turbine is associated with exactly one data point, and is colored according to the most likely regime it belongs to (b). The average incoming wind vector is denoted by an arrow. }
\label{fig:zones}
\end{figure}

Four distributions are identified in the data, corresponding to operational zones present in the wind farm (see Figure~\ref{fig:zones}). In the figure, the orange and purple cluster were extracted after applying the method a second time on the comprising subset of turbines; therefore, their turbines are expected to exhibit more similar behavior than within other zones. Although the upstream turbines are clustered together higher on the power curve, downstream turbines are located significantly lower. This is expected under the observed wind vector, as the wake effect becomes significantly more visible for incoming wind of less than 10 m/s \cite{wake_2011}.

To establish the link between the extracted operational zones and load profiles, the load duration distribution (LDD) and load revolution distribution (LRD) are analyzed, which are often used in analyses because of their standardization in turbine design \cite{iso_standard_design_2012}. The LDD/LRD are frequency distributions constructed by binning the power production range and counting the number of seconds/revolutions per bin. These distributions are constructed using the 1-second SCADA measurements of the power production and rotor speed during the 2-min interval.

\begin{figure}[!hb]
\begin{center}
\includegraphics[scale=0.8]{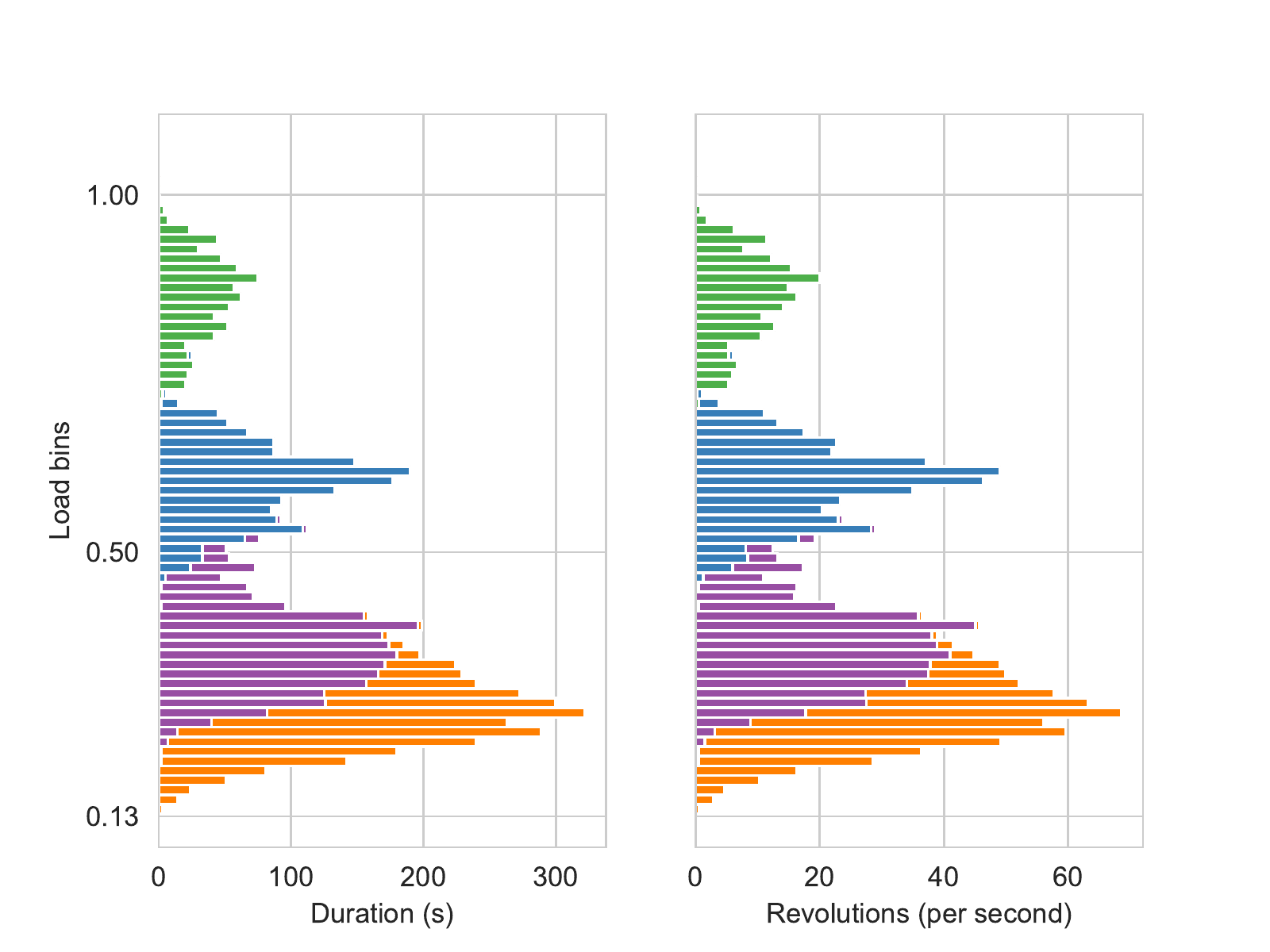}
\end{center}
\caption{Load duration (left) and revolution (right) distribution over all turbines in the wind farm. The measurements are colored according to the corresponding operational zones in Figure~\ref{fig:zones}. The loads are normalized between 0 and 1, respectively, denoting the minimum and maximum load possible.}
\label{fig:sharing_loads_twins}
\end{figure}

As shown in Figure~\ref{fig:sharing_loads_twins}, the detected zones can be directly related to load profiles, which can be described as four Gaussians. As mentioned before, the purple and orange zones consist of more similar turbines compared to other operational zones, which is reflected in the large overlap of the two respective load distributions.

These results show that compiling 1-second SCADA data between similarly behaving turbines provides accurate context descriptions on a short timescale. More specifically, the load profiles produce valuable insights into the load concentration and the lifetime consumption of the wind farm. This describes expected loading conditions and operational patterns in the farm under steady-state conditions. Deviations from these patterns are valuable indicators to the controller, as they could indicate the initiation of an event.

\section{Failure avoidance for rapid optimization}
\label{sec:failure_avoidance}

\begin{figure}[!hb]
\begin{center}
\includegraphics[scale=0.5,trim=0cm 7cm 5cm 0cm]{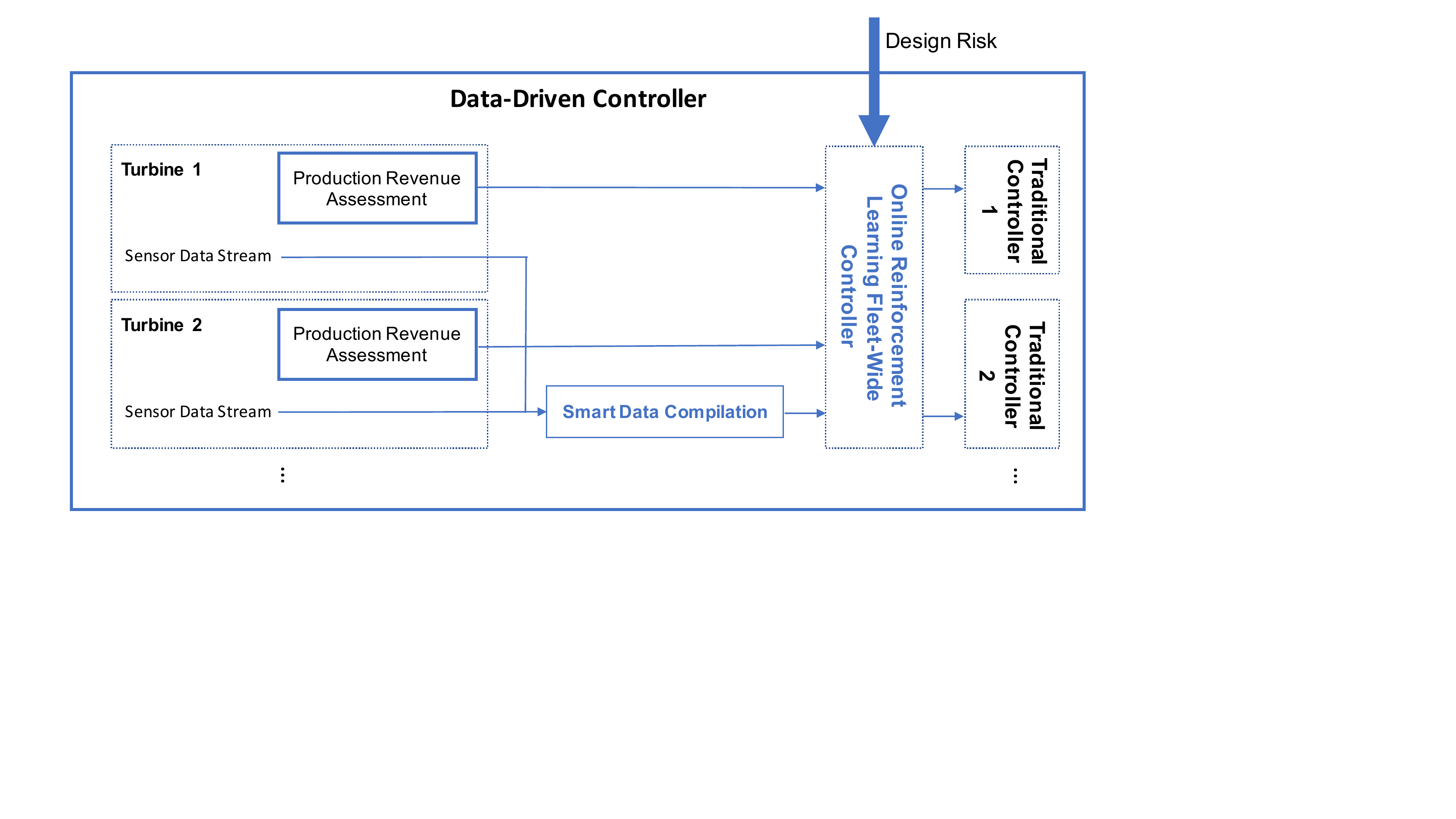}
\end{center}
\caption{Controller workflow for failure avoidance. The design risk is determined by the designer using the failure database, as depicted in Figure~\ref{fig:overview}, and used in the reward scheme of the RL-based fleetwide controller, together with the production revenues of the turbines. Statistically accurate environmental and operational conditions are provided by the ``Smart Data Compilation'' mechanism. The fleetwide RL controller then suggests an appropriate response to each of the turbines.}
\label{fig:controller_workflow}
\end{figure}

Reducing the number of failures occurring in the wind farm is the last step in our research approach, shown in Figure~\ref{fig:overview}. While highly focused research continues to further understand the failure mechanisms, the reliability of the existing fleet is optimized based on collected field data. Turbines already producing power in a fleet represent a huge initial investment. Making significant design modifications in-situ is nearly impossible. Therefore, this step optimizes the control strategy by \emph{learning} the connections between failure-driven events and turbine responses from fleet data. A data-driven top-level farm controller that continuously learns to coordinate the turbines is introduced, aiming to reduce the likelihood of encountering failure while promoting productivity (see Figure~\ref{fig:controller_workflow}). Depending on the designer's preference, a high or low cost can be associated with failures, allowing for conservative (i.e., fewer failures) or risky (i.e., higher productivity) control strategies.

Our approach is much the same as research linking smoking to cancer. While fundamental research investigates the biological intricacies of smoke and its potential carcinogenic properties, outside the laboratory the message is much more direct and simple. In general, to avoid cancer one should not smoke \cite{smoking_2004}. In our case, wind turbines have a central controller unit, which reads limited environmental data and then modulates variables, such as blade pitch and yaw, to harvest as much energy as possible while aiming to keep loading within predefined boundaries. In a similar way to humans avoiding to smoke, the controller will \emph{learn} from historical patterns to understand and eventually predict what combination of environmental events and controller response generally leads to maintenance events or failure. The controller will then be able to take action to avoid it entirely without explicitly understanding the mechanisms driving the failure. This concept is called \emph{failure avoidance} and describes a method for preventing failures.

A controller can only learn when it is allowed to move beyond expert knowledge and established protocols. \emph{Reinforcement learning} (RL) \cite{rl_2018} techniques are able to operate in real time and provide alternative exploratory control policies, while staying within the risk-free bounds established by the control designer. The controller learns to take decisions at the wind farm level, such that a cumulative reward linked to the consequences of all turbine responses is maximized. As this reward should be naturally linked to revenue, it can be defined in terms of power production and observed dynamic loads, used as a proxy for failure.

The control optimization technique used is REINFORCE (policy gradient) \cite{reinforce_1992}, which performs online optimization of policy parameters by sampling control decisions stochastically and evaluating alternative policies according to the defined reward scheme. Policy gradient has gained recent interest in numerous physical applications \cite{reinforce_robotics_2006, policy_grad_octopus_2014, rl_2018}. One of the reasons for this interest is the possibility of incorporating prior knowledge about the control strategy structure. For example, it makes sense to have a systematic shutdown of subsets of turbines when a storm cascades through the wind farm. Such knowledge can easily be incorporated in policy gradient algorithms. Additionally, policy gradient methods have the ability to deal with continuous data streams and control actions, in contrast to value-based RL methods \cite{rl_2018}.

Formally, the policy parameters $\theta$ are shifted toward the gradient-induced by their samples, with a magnitude of the cumulative reward until time $t$. 
\begin{equation}
\theta_{\text{new}} \leftarrow \theta_{\text{old}} + \alpha \nabla_\theta \log \pi_\theta (\text{context}_t) \sum^t_{i=1} R_i
\label{eq:reinforce}
\end{equation}
where $\alpha$ is the learning rate, and $\pi_\theta(\text{context}_t)$ is the control policy applied to the context (e.g., environmental factors, turbine conditions) observed at time $t$. This method is used because it is easy to place restrictions upon this stochasticity by using Gaussians to stay within the bounds of the established control protocols.

A controller can be specified for any event that may lead to failure. In order to know when a controller should be initiated, the associated event must be detected. This is done through the load profiles constructed in Section~\ref{sec:fleet_loads}, as they set the baseline for steady-state behavior. Discrepancies with respect to these profiles, together with other data sources (e.g., status logs), indicate what type of event is initiated.

The technique is illustrated on a case of excessive wind speeds going through the wind farm. As mentioned in Section~\ref{sec:gearbox}, dynamic loads on the gearbox induced by turbine responses, such as emergency stops, are suspected to be a major influence on the initiation of failure in the system. Detailed evidence of HSS bearing slip caused by grid-loss events is given. As both grid loss and emergency stops are torque-reversal events, emergency stops also have a big potential to induce slip \cite{grid_loss_2016}. This claim is supported by other studies as well \cite{slip_emergency_evidence_2013, slip_emergency_evidence_2015}. The slip is linked to the increased chance of bearing failure. Consequently, accurate prediction of these events allows for proactive preventive measures, increasing the bearing lifetimes over the entire wind farm.
Therefore, the focus of this work is on reducing the number of emergency stops during 1 hour of 1-second event logs during a known extreme storm event (i.e., the \emph{Santa Claus} storm of 2013) \cite{storm_2013}, while keeping turbines active as long as possible. Based on the wind measurements in the farm, the average wind vector is coming at an angle of $265.4^{\circ}$, which is almost orthogonal to the north-south rows of the wind farm. The turbine alarm logs are recorded each second: particularly the high-wind-speed ($> 25$ m/s) alarms and shutdowns of the turbines.\footnote{Data for turbines 05/5, 06/5, 07/5, and 08/5 are not available.}

\begin{figure}[!hb]
\begin{subfigure}{0.48\textwidth}
\begin{center}
\includegraphics[scale=0.68,trim=4cm 9.45cm 0cm 12cm]{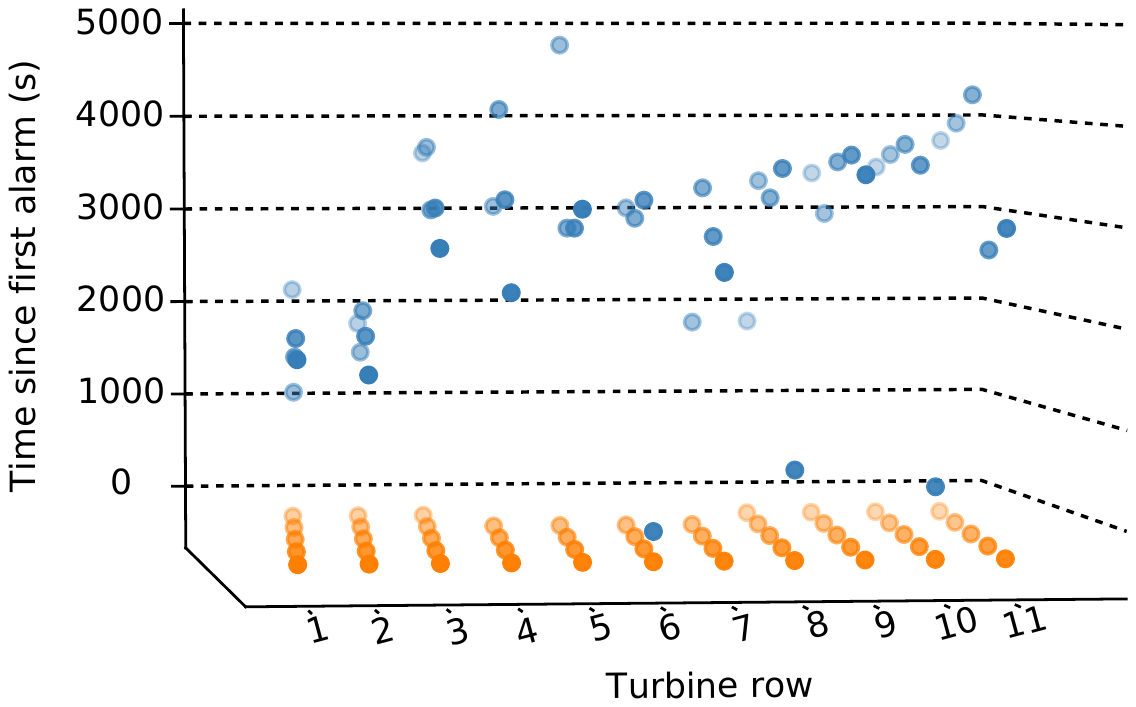}
\caption{High-wind-speed alarms}
\label{fig:shutdown_alarms}
\end{center}
\end{subfigure}
\begin{subfigure}{0.48\textwidth}
\begin{center}
\includegraphics[scale=0.45]{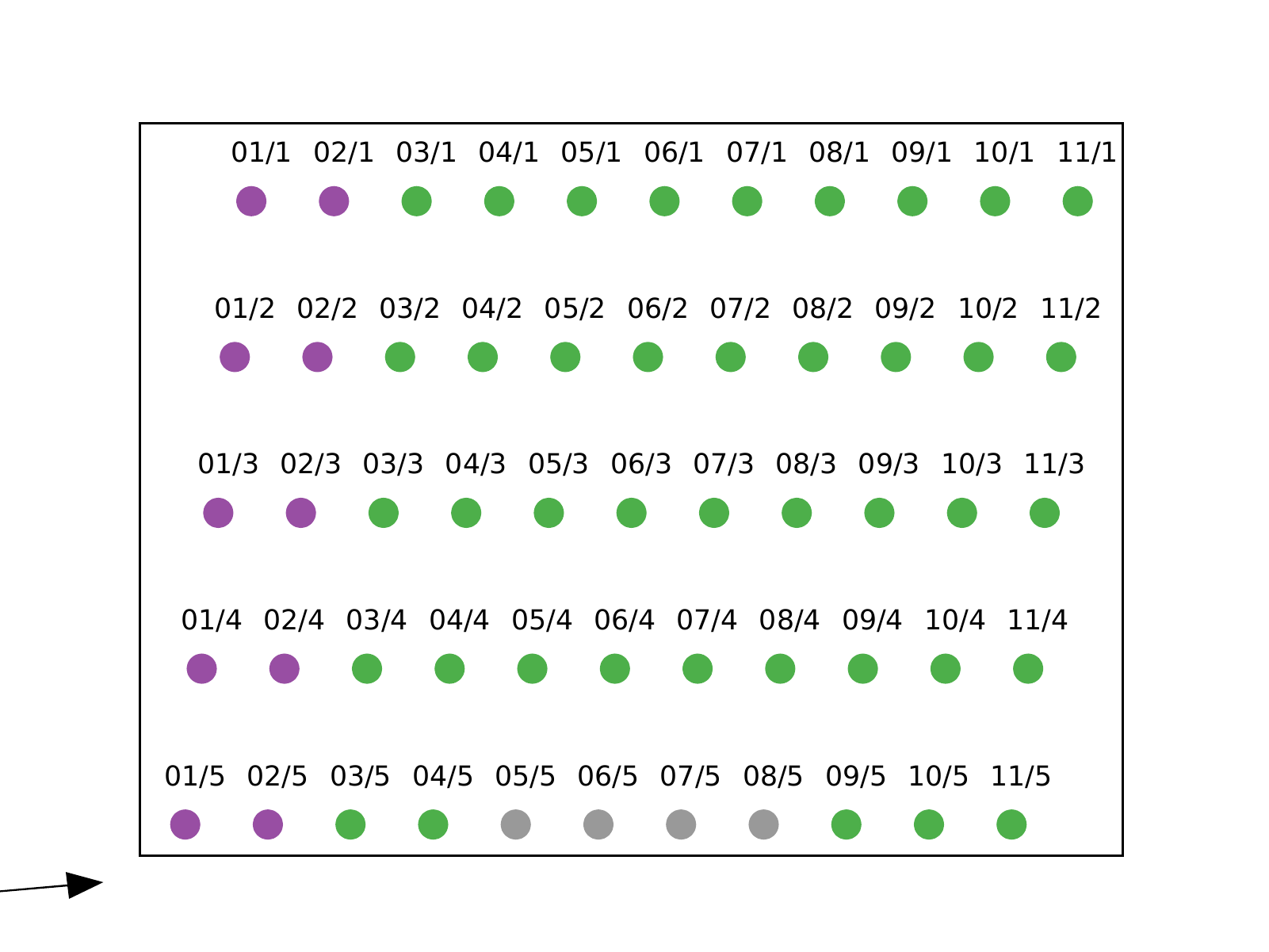}
\caption{Time stamp clustering}
\label{fig:shutdown_clustering}
\end{center}
\end{subfigure}
\caption{The time stamps of the initial event alarms are plotted for each turbine (a). These time stamps are clustered using GMMs into two groups (b). The incoming wind vector is denoted by an arrow.}
\label{fig:shutdown}
\end{figure}

Figure~\ref{fig:shutdown_alarms} shows when the event is first observed by each turbine. As there is a linear increasing trend over the rows, a preventive row-based mechanism is proposed, wherein each row waits the necessary amount of time $d_r$ after the previous row to perform a shutdown. Pragmatically speaking, this approach makes sense, as the wind will propagate through the farm and is influenced by the turbines it passes. Linked to this concept, it is interesting to note that the first two rows seem to trigger the alarm significantly earlier than the rest of the wind farm. Moreover, it is interesting to note that the first two rows seem to trigger the alarm significantly earlier than the rest of the farm. By inspecting the load profiles of the front turbines, discrepancies should be seen and the high-wind-speed event should be detected based on the high-wind-speed alarms. As no operational data are available to construct the load profiles, the GMMs are instead applied to the time stamps to verify the observed discrepancies, and smooth the clustering by taking the most occurring label among the eight surrounding neighbors (see Figure~\ref{fig:shutdown_clustering}).
Embedding this prior knowledge into the control protocol presents an opportunity to use the upstream rows as beacons, signaling the event to the farm controller. Once signaled, a preventive control strategy can be initiated, such that the downstream rows keep producing until a safe shutdown can be carried out.
The event is assumed to be started when three turbines of the two beacon rows signal the alarm.

For the downstream rows, indexed in order from west to east by $r \in \{3, \dots, 11\}$, the delays $d_r$ of the row-based shutdown policy are optimized using REINFORCE. 
In order to encourage exploration, Gaussians are placed on the delays with a standard deviation of 30 seconds, such that about $95\%$ of the samples fall within 1 minute from the current best delays. Thus, in accordance with Equation~\ref{eq:reinforce}, $\pi_\theta$ is the row-based mechanism, and the context given to the function $\pi_\theta$ is the current row index $r$, which then returns a control decision, namely the time until a shutdown is initiated. The resulting policy is given by
\begin{equation}
d_r(i) \sim \mathcal{N}(\theta_r, 0.99^i \times 30)
\label{eq:delay}
\end{equation}
where possible actions $d_r$ at iteration $i$ are explored by sampling from a Gaussian with a row-dependent mean and standard deviation 30 decaying over time with a factor of 0.99. The standard deviation of 30 seconds is chosen such that a delay within 1 minute of the mean is sampled 95\% of the time (confidence interval). The decay factor 0.99 allows for a sufficient amount of exploration before the control strategy converges. The number of iterations is set to 1000 to ensure convergence of the control strategy.

As continuous activity of turbines results in a higher profit margin, it is important to keep the rows operational for as long as possible without inducing significant loads on the gearbox by emergency braking.
Preferably, the reward scheme should reflect the total revenue gained by selecting certain control decisions over others, and therefore depends on the power production and observed loads of each turbine at each time step during the event.
Although it is feasible to approximate these quantities from real-time fleet data, current test beds do not necessarily support the verification of alternative control strategies in-situ. A simplified reward scheme that encourages production time is used while giving a penalty $P$ for being active during the storm. Thus, if a turbine in row $r$ observes the storm at time $T_{storm}$, then the reward given to the turbine at time $t$ is:

\begin{equation*}
\begin{split}
R(t, T_{storm}, \{d_i\}^r_{i=3}) =
\begin{cases}
+1 \text{\ \ if } (t < T_{storm}) \text{ and } (t \le \sum^r_{i=3} d_i)\\
-P \text{ if } (t \ge T_{storm}) \text{ and } (t \le \sum^r_{i=3} d_i)\\
0 \text{\quad\ otherwise}
\end{cases}
\end{split}
\end{equation*}

\noindent
Optimizing the cumulative reward through this proxy function validates alternative control strategies suggested by the reinforcement learner, and provides a proof of concept for the data-driven control mechanism. Future work includes extending the reward scheme using real load indicators to express the lifetime consumption of the turbines.

\begin{figure}[!hb]
\begin{center}
\includegraphics[scale=0.6]{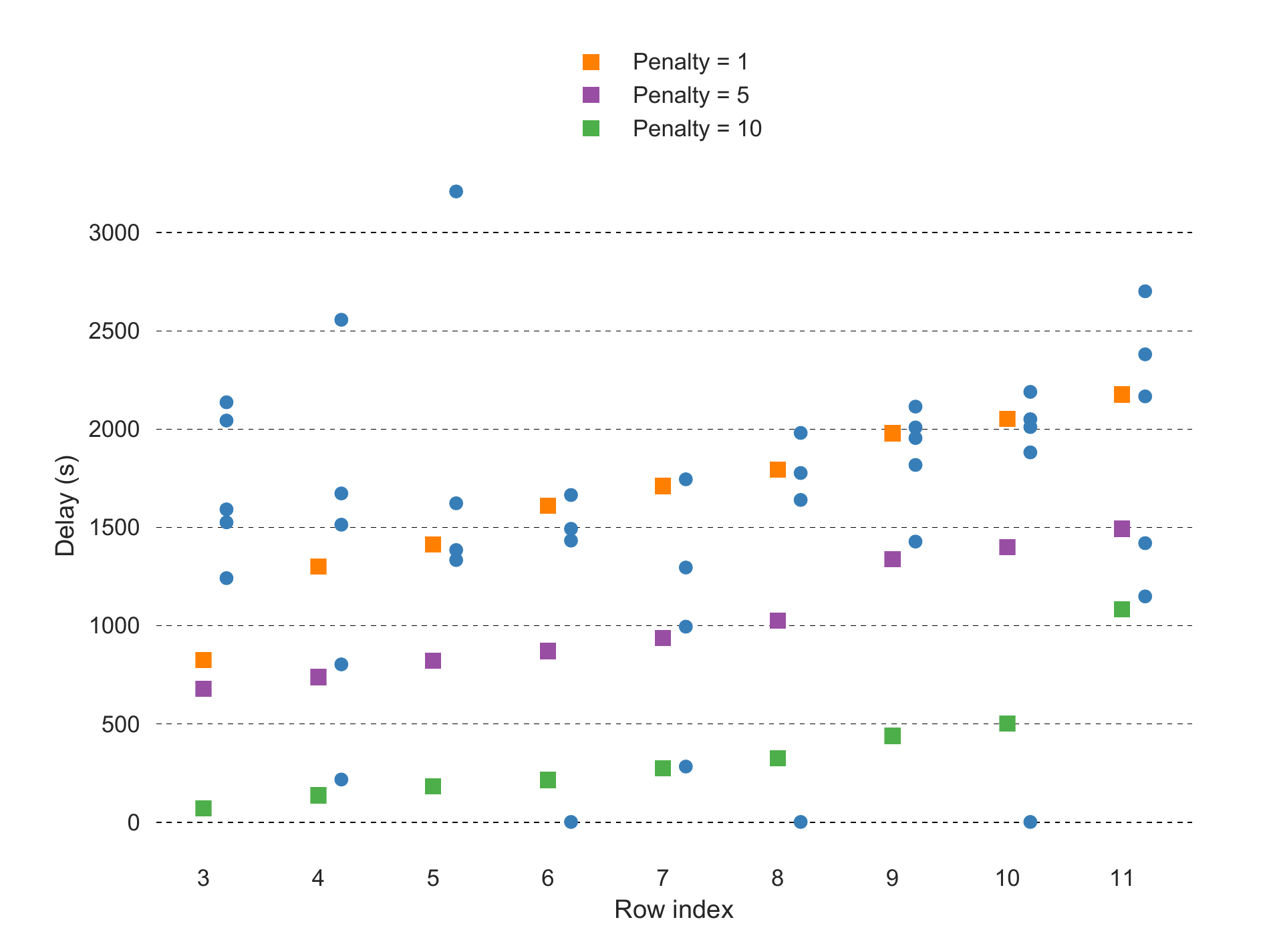}
\end{center}
\caption{The learned control policies for various penalties, ranging from conservative ($P = 10$) to risky ($P = 1$) strategies. The high-wind-speed alarms are depicted per row by blue dots.}
\label{fig:rl_policies}
\end{figure}

Figure~\ref{fig:rl_policies} shows the optimized delays for different wind farm operation strategies during the storm. Convergence tests are carried out to verify whether REINFORCE found the optimal control strategy with respect to the given reward function. All three control strategies passed those tests. As shown, there is a linear trend over the wind farm to which shutdowns should be carried out. This trend is in accordance with the data, in which the detected event occurs only a few minutes after the proposed shutdowns. Increasing the penalties creates more conservative strategies, allowing more alarms to occur within one row before a shutdown. This is useful to dampen any response to alarms caused by small wind gusts.

\section{Discussion}

The operational zones are extracted by fitting GMMs on the given fleet data. Bayesian GMMs are statistical models that aim to find the mixture of Gaussians that best represents the data while limiting the number of Gaussians used. It should be clarified that, although the operational parameters may not be separable into a set of Gaussians, they can still be well-approximated by our model. This is because GMMs are dense in the set of all probability distributions \cite{gaussian_mixture_dense_2010}, which means that any underlying data distribution can be approximated arbitrarily close using a set of Gaussians.

To increase confidence in the validity of this reasoning with respect to operational zones, each of the found clusters per operational parameter should pass an independent Shapiro-Wilk normality test. This is indeed the case for most clusters, as the normality null hypothesis cannot be rejected (p-values $> 0.05$), except for the rotor speed of the blue cluster. When examining Figure~\ref{fig:zones}, this makes sense, as the blue points are skewed at the rated output speed. An expert may now decide to further decompose the blue cluster into two Gaussians, similar to the hierarchical subclustering of the orange and purple clusters in Section~\ref{sec:fleet_loads}. As the blue cluster resulted in a subcluster with only three turbines, it is kept as is.
The normality on a 1-second basis can also be verified visually in Figure~\ref{fig:sharing_loads_twins}, wherein the load profiles are Gaussian shaped, even though the full data distribution is multimodal. Once the normality is verified and the operational zones are established, the average statistics for the load profiles in Section~\ref{sec:fleet_loads} are well-grounded, as every data point is considered to be a noisy measurement from the mean of the respective Gaussian.

It should be noted that, while the clustering using GMMs can be considered to be objective, the choice of imposing a hierarchy is subjectively made by the designer. Nevertheless, such a hierarchy provides interesting insights, as it gives a clear overview of the load spectra present in the farm at different levels of detail. For example, the results in Figure~\ref{fig:sharing_loads_twins} show how the LDD/LRD for the first-level clusters (e.g., green and blue) are better separated than the second-level clusters (e.g., orange and purple), even though Gaussians can be clearly distinguished for all clusters.

Failure avoidance is formalized as a reinforcement learning problem. Reinforcement learning is different from other \emph{supervised learning} predictors, as it does not solely rely on direct information about suitable control strategies. Such information is only partially available through thorough analysis of dynamic loads in test rigs by using the aforementioned methodological steps. Instead, RL aims to find the optimal control policy through targeted exploration of suitable actions in the field. Because of its model-free nature, it is capable of optimizing highly complicated cost functions without fully relying on in-depth domain knowledge about the links between the dynamic event and failure. The other alternative is the use of planning techniques, which require a full predefined model of the wind farm dynamics \cite{rl_2018}; however, this type of model is not available in our setting. Naturally, additional knowledge of the load spectra guides the learning process, resulting in more versatile and responsive control policies. This combination of physical insights into the farm dynamics and data-enabled control optimization is fundamental to the research strategy depicted in Figure~\ref{fig:overview} for maximum reliability improvement.

REINFORCE is used for the failure avoidance mechanism. This method is a learner for data-driven control and is able to optimize complex and nonanalytic reward functions. It must be noted that REINFORCE converges to a control strategy that is locally optimal in terms of the cumulative reward, because of the iterative updates of the control parameters \cite{rl_2018}. In many physical applications, these local optimas still perform well \cite{reinforce_robotics_2006, policy_grad_octopus_2014}. The control designer may add additional exploration to the control policy to prevent getting stuck in local optimas (e.g., by increasing the standard deviation in Equation~\ref{eq:delay}).

The findings in this work show the potential of our a novel multitiered methodology and data-driven methods aimed at reliability optimization of wind turbines and positively impacting the future design of turbine components. Researched findings from a mechanism level are linked to findings in both highly instrumented and real-world farm data streams. By applying this methodology to more suspect load cases, clear advantages can be realized in the development of control strategies, as well as insights gained in nonstraightforward failure modes.

First, better indicators are provided to researchers and designers to fuel fundamental insights and influence more optimal future designs. This is accomplished by identifying operational zones in the wind farm and linking them to loading conditions. Additionally, the failure avoidance control mechanism can be used to routinely upgrade turbine software in the field such that the wind turbines are less prone to risky behavior even before designers understand the fundamental causes of all failure modes.

The data-driven model is presented as a proof of concept and is validated in a qualitative study on high-wind-speed stops, outlining the mechanism for event detection and subsequent farmwide control. However, because of the limited availability of field data, a thorough quantitative study on an abundant variety of cases is not possible. This sets the main limitation of the model, as the evaluation of the failure avoidance method is restricted to off-line control learning, which obfuscates the online decision support capabilities of the RL method.

Nevertheless, this work shows that the outlined approach has great potential in the context of failure analysis and avoidance. It provides a mechanism for control designers to take into account the dynamic loads that drive failure. This is an improvement upon state-of-the-art wind farm control research, which mainly focuses on static loads and power production. The framework is generic enough to be applied to several cases. For example, more turbines are fitted with high wind ride-through mechanisms \cite{ride_through_2018}, which essentially allows the power curve to decrease smoothly at high wind speeds, rather than abruptly stop through emergency brakes. The rate at which the power curve drops can be optimized by the controller, as it should aim to keep the turbines operational as long as possible, but at the same time reduce the penalties given for staying operational during dynamic events. This is a setting that is similar to the demonstration case presented in Section~\ref{sec:failure_avoidance}. Additionally, although this work focuses on event-driven dynamic loads, the framework can be applied in steady-state environmental conditions as well. For example, yaw behavior or cut-in/-off can be controlled in order to optimize complex nonlinear health statistics of the turbines, such that unhealthy turbines with a high load profile (for specific wind conditions) are less operational to improve reliability.

\section{Conclusion}

This work presented a novel multitiered framework for investigating the links between dynamic loads and failures in the field, and introducing them into targeted wind farm control strategies. Additional evidence has been provided for gearbox failure caused by grid-loss events. A load-profiling mechanism based on operational farm dynamics was introduced. This mechanism provides high-level patterns about the load spectra occurring in the wind farm, and can be used as a baseline for the farm controller. Finally, a data-driven farm controller was introduced to prevent emergency stops, which are known to induce a significant amount of load and can lead to failure. The demonstration shows how domain knowledge about dynamic loads and failure can be incorporated in wind farm control, which is crucial to improve reliability within the wind farm. The inclusion of failure-based objectives in wind farm control strategies is a novel step in the field of turbine control design and toward reliability improvement of wind turbines.

In future work, the load-profiling and failure-avoidance methods will be extended to yield a broader spectrum of turbine operational events and thus more statistically representative outcomes. Additionally, an event detection mechanism will be developed to properly detect discrepancies in the load profiles and to recognize which event is taking place. This will be done through the means of supervised machine learning techniques based on historic data of several events.
The investigation on dynamic event-driven loads and related failures will continue to further demonstrate the potential of the proposed methodology. The failure avoidance mechanism requires validation on real-time experimental test beds to show the full potential of the online decision support capacities of the REINFORCE method. Those decisions will be compared to current wind farm control strategies and help pinpoint the potential for improvements in productivity and reduction of dynamic loads.
The evolution and further development of these algorithms and the subsequent development of a deeper real-time database will enable an increase in system understanding and targeted control choices to optimize energy production and infrastructure investment. 

\section{Acknowledgments}

The authors would like to acknowledge FWO (Fonds Wetenschappelijk Onderzoek) for their support through the SB grant of Timothy Verstraeten (1S47617N). Additionally, the authors would like to acknowledge the financial support of VLAIO (Agentschap Innoveren \& Ondernemen) through the SBO project HYMOP (Iwt150033). Furthermore, the authors thank their partners for delivering the monitoring data. The authors thank the anonymous reviewers whose comments/suggestions helped improve and clarify this manuscript. Finally, thanks to Sheri Anstedt (NREL) and Pieter J.K. Libin (VUB) for proof reading the manuscript.

This work was authored [in part] by the National Renewable Energy Laboratory, operated by Alliance for Sustainable Energy, LLC, for the U.S. Department of Energy (DOE) under Contract No. DE-AC36-08GO28308. Funding provided by the U.S. Department of Energy Office of Energy Efficiency and Renewable Energy Wind Energy Technologies Office. The views expressed in the article do not necessarily represent the views of the DOE  or the U.S. Government.  The U.S. Government retains and the publisher, by accepting the article for publication, acknowledges that the U.S. Government retains a nonexclusive, paid-up, irrevocable, worldwide license to publish or reproduce the published form of this work, or allow others to do so, for U.S. Government purposes.





\section{References}

\bibliographystyle{model1-num-names}
\bibliography{refs.bib}







\end{document}